\documentclass[11pt,]{article}
\usepackage{lmodern}
\usepackage{amssymb,amsmath}
\usepackage{ifxetex,ifluatex}
\usepackage{verbatim}
\usepackage{lscape}
\usepackage{fixltx2e} % provides \textsubscript
\ifnum 0\ifxetex 1\fi\ifluatex 1\fi=0 % if pdftex
  \usepackage[T1]{fontenc}
  \usepackage[utf8]{inputenc}
\else % if luatex or xelatex
  \ifxetex
    \usepackage{mathspec}
  \else
    \usepackage{fontspec}
  \fi
  \defaultfontfeatures{Ligatures=TeX,Scale=MatchLowercase}
\fi
\IfFileExists{upquote.sty}{\usepackage{upquote}}{}
\IfFileExists{microtype.sty}{%
\usepackage{microtype}
\UseMicrotypeSet[protrusion]{basicmath} % disable protrusion for tt fonts
}{}
\usepackage[margin=1in]{geometry}
\usepackage{hyperref}
\hypersetup{unicode=true,
            pdftitle={M-models},
            pdfauthor={Gonzalo},
            pdfborder={0 0 0},
            breaklinks=true}
\usepackage{graphicx,grffile}
\makeatletter
\def\maxwidth{\ifdim\Gin@nat@width>\linewidth\linewidth\else\Gin@nat@width\fi}
\def\maxheight{\ifdim\Gin@nat@height>\textheight\textheight\else\Gin@nat@height\fi}
\makeatother
\setkeys{Gin}{width=\maxwidth,height=\maxheight,keepaspectratio}
\IfFileExists{parskip.sty}{%
\usepackage{parskip}
}{% else
\setlength{\parindent}{0pt}
\setlength{\parskip}{6pt plus 2pt minus 1pt}
}
\ifx\paragraph\undefined\else
\let\oldparagraph\paragraph
\renewcommand{\paragraph}[1]{\oldparagraph{#1}\mbox{}}
\fi
\ifx\subparagraph\undefined\else
\let\oldsubparagraph\subparagraph
\renewcommand{\subparagraph}[1]{\oldsubparagraph{#1}\mbox{}}
\fi

\let\rmarkdownfootnote\footnote%
\def\footnote{\protect\rmarkdownfootnote}

\usepackage{titling}

\usepackage{float}
\usepackage[svgnames]{xcolor}
\usepackage{booktabs, multirow}
\usepackage{rotating}
\usepackage{booktabs}
\usepackage[font=small,labelfont=bf]{caption}

\usepackage{makecell} %% dividir celdas
\usepackage{mathdots}

\usepackage[round]{natbib}
\def\Q{\boldsymbol{Q}}

\def\1{\bf 1}
\def\0{\bf 0}

\newfont{\Sc}{eusm10}

\def\OOmega{\mbox{\boldmath{$\Omega$}}}

\title{A fast approach for analyzing spatio-temporal patterns in ischemic heart disease mortality across US counties (1999-2021)}
\date{}
\author{Urdangarin, A.$^{1,2}$, Goicoa, T.$^{1,2}$, Congdon, P.$^{3}$, Ugarte, M.D.$^{1,2*}$\\
\small {\textit{$^1$ Department of Statistics, Computer Science, and Mathematics, Public University of Navarre, Pamplona, Spain.}} \\
\small {\textit{$^2$ INAMAT$^2$, Public University of Navarre, Pamplona, Spain.}} \\
\small {\textit{$^3$ School of Geography, Queen Mary University of London, London, UK.}} \\
\small {$*$Correspondence to Mar\'ia Dolores Ugarte, Department of Statistics, Computer Science, and Mathematics,} \\
\small {Public University of Navarre, Campus de Arrosadia, 31006 Pamplona, Spain.} \\
\small {\textbf{E-mail}: lola@unavarra.es }}

\usepackage{listings}
\usepackage{caption}
\usepackage{subcaption}

\begin{document}
\maketitle

%%---------------------------------------------------------------------------------------
%% ABSTRACT
%%---------------------------------------------------------------------------------------
\begin{abstract}
Ischaemic heart disease (IHD) remains the primary cause of mortality in the US.   This study focuses on using spatio-temporal disease mapping models to explore the temporal trends of IHD at the county level from 1999 to 2021. To manage the computational burden arising from the high-dimensional data, we employ scalable Bayesian  models using a "divide and conquer'' strategy.  This approach allows for fast model fitting and serves as an efficient procedure for screening spatio-temporal patterns.
 Additionally, we analyze trends in four regional subdivisions, West, Midwest, South and Northeast, and in urban and rural areas. The dataset on IHD contains missing data, and we propose a procedure to impute the omitted information. The results show a slowdown in the decrease of IHD mortality in the US after 2014 with a slight increase noted after 2019. However, differences exists among the counties, the four big geographical regions, and rural and urban areas.

\end{abstract}


%%%%%%%%%%%%%%%%
% INTRODUCTION %
%%%%%%%%%%%%%%%%

\section{Introduction}

In the 1960s, the United States experienced a peak in deaths related to ischaemic heart disease (IHD). Subsequently, mortality rates associated with IHD started to decline. This positive trend can be attributed to advances in diagnosis and treatment, coupled with the adoption of healthier lifestyles. Factors contributing to the decline include a reduction in smoking rates, improvements in managing blood pressure, and the control of blood cholesterol levels \citep{Ford2007}. By the year 2000, IHD mortality rates were about one third lower than their baseline in the 1960s \citep{Mensah2017}. However, despite these improvements, the IHD continues being the leading cause of mortality in the US.

From 2010 onward, a flattening is observed in IHD mortality rates. The slowdown in the decline of IHD mortality rates has been argued to be the primary factor contributing to the stagnation in life expectancy growth in the US after 2010 and is particularly notable in contrast to the ongoing decrease in mortality rates observed in other peer high-income countries \citep{Mehta2020}. Some studies also comment on a possible connection between the deceleration in the decline of heart disease mortality and the increase in the prevalence of obesity and diabetes \citep[see, for example][]{Wilmot2015, Vaughan2017, Vaughan2018}.

The overall mortality rates of US may not capture the variations among different geographic regions and demographic groups  and several studies have highlighted varying trends in IHD mortality among geographic regions \citep{Cooper2000, Kulshreshtha2014}. Moreover, considering trends from 1999 to 2009,
\cite{Kulshreshtha2014} report that favorable US trends in IHD mortality conceal disparities among some demographic groups (e.g., non-metropolitan areas). Nevertheless, such studies of IHD mortality trends do not employ a formal statistical modelling approach. Recently, \cite{Quick2018} present a non-separable multivariate spatio-temporal Bayesian model (MSTCAR) to gain knowledge about how heart disease mortality rates at county level in the US evolve in time for different sex-race groups. The model allows for the estimation of correlations between sex-race groups in each year of the study period (1973-2010).
%{\color{red} The model allows for the estimation of between sex-race group correlations in every year of the study period (1973-2010).
The same model has been used  to study stroke mortality among old age groups (65+) at the county level in United States during the period 1973-2013 \citep{Quick2017}, and for middle-age adults (35-64) and adults (65+) in the period 2010-2016 \citep{Hall2019}.
Without being exhaustive, other applications of the MSTCAR model include examining the ranking of heart disease mortality rates at the county level by race-sex groups during the period 1973-2015 (Vaughan et al., 2019), as well as studying recent trends in coronary heart disease mortality by sex, race, and age group at the county level (Vaughan et al., 2020). The MSTCAR model used in these studies offers valuable insights into potential drivers of shifts in heart disease mortality. However, it also has some disadvantages. A key one is that it relies on computationally intensive MCMC methods \citep[see for example][]{Quick2019}. The computation burden may be huge as the number of hyperparameters increases dramatically with the socio-demographic groups and the time periods leading to potential convergence problems. As an example, \cite{Quick2017} has 41 periods of time and 3 age groups leading to $246$ between-groups variance-covariance hyperparameters plus 41 additional temporal correlation hyperparameters. Moreover, the number of counties in this study is 3099 leading to the estimation of about 400000 rates parameters. A second drawback is the use of a global spatial pattern for the whole US that might not be the best option if different amount of smoothing is required depending on the local regions.

In this work, we consider simpler but highly computationally efficient spatio-temporal models within a hierarchical Bayesian framework. We use the models to describe the spatial distribution of IHD mortality in the US and illustrate its evolution over the period 1999-2021. In particular, we examine temporal trends of IHD mortality risks for the whole US, within each of its four geographic regions —West, Midwest, South, and Northeast— and at the county level to identify disparities. The study also describes the variations in the temporal trends between rural and urban areas. Additionally, and by a mere visual inspection of the trends, we pursue identifying change points, that is, instances where a change might occur in the prolonged declines, for example, a flattening or increasing of the trends after a certain point \citep{berrett2021, Murphy2022}. %\textcolor{blue}{Due to the rising number of missing values in the data, our analysis does not address disparities among demographic groups such as age, race, or gender.}
Changes in the geographical patterns of a disease over time often involve nonlinear trends that are typically best analyzed using flexible random effects that account for temporal autocorrelation. Moreover, spatio-temporal models will account for variations in temporal trends across different counties, incorporating spatio-temporal interaction terms.  The models are very appealing from an interpretative standpoint, as they include a spatial effect that can be related to an underlying spatial risk throughout the period, a temporal effect that captures the overall trend of the entire study region, and an interaction term that can be viewed as either the deviation of county-level temporal evolution from the overall global trend or the deviation of each year's spatial pattern from the underlying spatial risk surface. Selection of optimal interaction schemes in relatively large datasets, as here, can be computationally challenging, and many studies may resort to default interaction assumptions. An additional challenge of the IHD data is the notable presence of missing values, i.e suppressed data, since we use publicly available IHD data provided by National Center for Health Statistics (NCHS) through the Center for Disease Control and Prevention (CDC), and the Wide-ranging Online Data for Epidemiologic Research (WONDER), and counts less than 10 are not provided for confidentiality reasons.
Hence, before conducting the spatio-temporal analysis of the data, suppressed counts have been imputed, with imputation based on spatial borrowing of strength by fitting spatial models to each year of the study period. Using Bayesian spatial models for highly censored data from CDC WONDER has been recently recommended \citep{Quick2019}. To handle the large size of the dataset (US counties are analyzed over a relatively long time span of 23 periods), we employ scalable Bayesian spatio-temporal models using a \lq\lq divide and conquer'' approach,  adapted to computation with large datasets \citep[see][]{OrozcoAcosta2023, Heaton2019}.  The \lq\lq divide and conquer approach'' splits the whole region into smaller subdomains and fits spatio-temporal models in each subregion. Hence, it overcomes computational challenges associated to large datasets in Bayesian inference techniques, including simulation based MCMC algorithms or integrated nested Laplace approximations (INLA) in spatio-temporal models with complex spatio-temporal interaction terms. Additionally, the key point is that the approach we use allows for inducing different degree of smoothing across the areal units within each subdomain and mitigates border effects by incorporating neighboring areas from adjacent regions. As a result, this \lq\lq divide and conquer'' approach is flexible, simple to understand, and improves model fitting compared to a single model for the whole US. The latter assumes a kind of data stationarity across the entire map, an assumption that is less likely to be valid in large spatial domains.

The rest of the paper is organized as follows. Section 2 describes the data, the imputation process for missing data and the spatio-temporal models. In Section 3 we present the results, and finally, the paper closes with a discussion.

%%%%%%%%%%%%%%%%%%%%%%%%
% MATERIAL AND METHODS %
%%%%%%%%%%%%%%%%%%%%%%%%

\section{Material and methods}

\subsection{Ischaemic heart disease mortality data in United States}

The IHD annual mortality totals for each county in the US between 1999 and 2021, along with the corresponding county populations, has been obtained from the CDC-WONDER web site ({\url{https://wonder.cdc.gov/}).  Here, we use the publicly available data provided by NCHS to CDC-WONDER with counts less than 10 suppressed by confidentiality reasons, but access to the unrestricted dataset may be possible by submitting a proposal to NCHS subject to certain conditions difficult to meet for non-US citizens. We denote as IHD deaths those with cause codes I20-I25 (ischaemic heart disease) in the International Classification of Diseases Tenth Revision (ICD-10).

The data set includes 3105 counties from all the US states except Alaska and Hawaii. Additionally, Puerto Rico, Samoa, the Dukes and Nantucket islands in Massachusetts, as well as San Juan Island in Washington, are excluded. Details regarding the data preprocessing procedures are provided in Supplementary Material A. The US is administratively divided into four geographic regions, namely, West, Midwest, South, and Northeast, encompassing the selected 48 states. Additionally, following \cite{Kulshreshtha2014}, we will classify counties into three population levels:  large metropolitan areas or metros, medium metros, and non metros or rural areas. More precisely, large metros include counties with more than 1 million people; medium metros include counties with a population ranging from 50,000 to 999,999 people, and non metros or rural areas comprise counties with less than 50,000 people.

Figure \ref{fig:US_division} (top) illustrates the geographic division and demographic classification of the US in 2021. Large and medium metros constitute $1.3\%$ and $29.9\%$ of the counties respectively whereas rural metros account for $68.8\%$ of the counties. The majority of urban counties are located along the coastal borders of the US. Figure \ref{fig:US_division} (bottom) shows the division of the US in four big regions, West, Midwest, South and Northeast. Among these four geographic regions, the Northeast is the only one that contains more urban counties than rural ones.

\begin{figure}[htbp]
	\centering
	\includegraphics[width=0.8\textwidth]{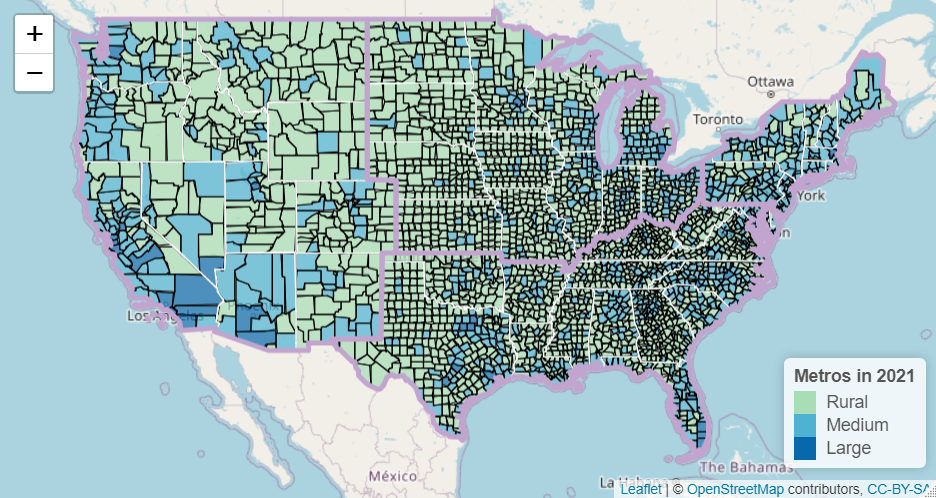}\\\vspace{0.2cm}
	\includegraphics[width=0.8\textwidth]{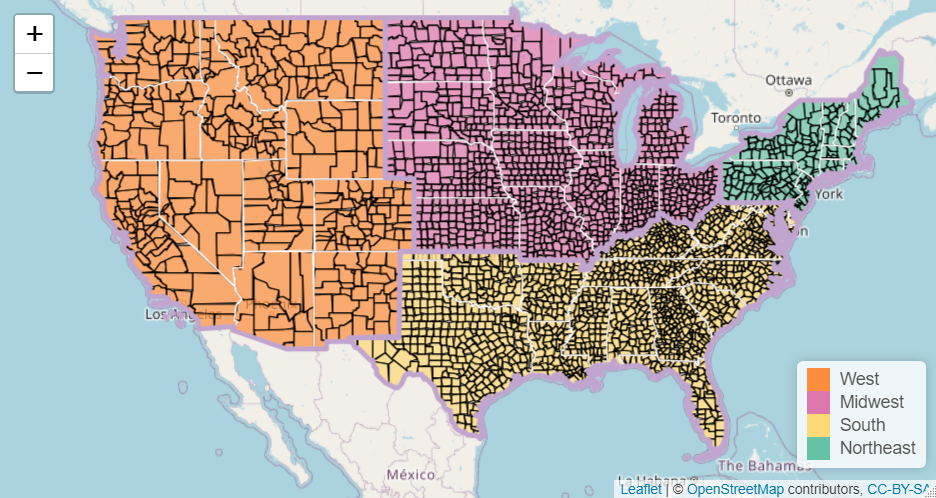}
	\caption{Top figure displays the geographical division of the US: black lines define counties, white lines represent state boundaries and purple lines indicate the division into four geographic regions (West, Midwest, South, and Northeast). The color-filled representation within counties corresponds to population levels (green for Rural metros, light blue for Medium metros and dark blue for Large metros). Bottom figure shows in different color the four big geographical regions of the US.}
	\label{fig:US_division}
\end{figure}

 Due to confidentiality issues,  $12.5\%$ of total observations are missing. To be more specific, $3.7\%$ of counties exhibit missing counts for each year in the data set.  Approximately $6.3\%$ of counties have between 15 and 22 years with missing counts, while around $15.2\%$ of counties have less than 15 years with missing data. Most of the counties with missing counts are rural areas located in the West and Midwest regions of US.  More precisely, 41.4\% of the counties in the West, 30.6\% of counties in the Midwest, 19.8\% of counties in the South and 2.8\% counties in the Northeast have missing values at least one year of the period. Figure B.1 and Table B.1 in the Supplementary Material B show the distribution of the counties with missing values across the US and the percentage of available and missing counts respectively.

\subsection{Imputation of missing counts}

Before proceeding with the spatio-temporal analysis of the IHD mortality data, missing data have been imputed. The imputation procedure involves fitting a spatial disease mapping model for each year of the period to estimate the relative risks of the counties with missing values. The missing counts are then replaced with the predicted values obtained from the fitted model.  \cite{Quick2017} comment on the convenience of using Bayesian spatial models as a way to provide synthetic data replacing the suppressed counts.

Spatial disease mapping models borrow strength over neighbouring units in space, thereby reducing the variability in disease risk estimates.  This is particularly advantageous when studying rare diseases or areas with small populations, which aligns with the characteristics of the counties with missing values,  so imputation is based on spatial borrowing strength. Various proposals are available for the spatial components of the model. In disease mapping, conditional autoregressive (CAR) \citep{Besag1974} priors on the spatial random effects are widely adopted to address spatial dependence in the data.

Let us denote as $Y_{it}$ the number of deaths due to IHD in the $i$th county ($i=1, \dots, S$) and year $t$ ($t=1, \dots, T$). Specifically, the dataset consists of $S=3105$ counties and spans $T=23$ time periods, covering the years from 1999 to 2021. In this study we consider $T$ independent spatial models with the following formulation of each time period
\begin{eqnarray}\label{eq:spatial_model}
	Y_{it} \arrowvert r_{it} &\sim& Poisson(e_{it}r_{it}) \\\nonumber
   \log r_{it} &=& \alpha_{0}^{t}+\xi_{i}^{t},
\end{eqnarray}
where $r_{it}$ and $e_{it}$ are the relative risks and the expected cases of the $i$th county at year $t$, respectively. The expected cases are computed as $e_{it}=N_{it} m_{t}$ where $N_{it}$ is the population at risk in the $i$th county at year $t$ and $m_{t}$ is the overall rate for each year $t$ computed as $m_{t}=\dfrac{\sum_{i=1}^{S} Y_{it}}{\sum_{i=1}^{S} N_{it}}$. The intercept $\alpha_{0}^{t}$ can be interpreted as an overall risk in year $t$ and $\boldsymbol{\xi}^{t}=(\xi_{1}^{t}, \xi_{2}^{t}, \dots, \xi_{S}^{t})^{'}$ is the vector of spatial random effects of the spatial model for year $t$ that follows a BYM2 prior \citep*{Riebler2016}. More precisely, $\boldsymbol{\xi}^{t} \sim N(\mathbf{0}, (\sigma_{\xi}^{t})^2 \OOmega^{t}_{\xi})$, where the covariance matrix $\OOmega^{t}_{\xi}$ is defined as a sum of a spatially structured term, determined by a spatial adjacency matrix $\boldsymbol{Q}_{\xi}$, and a spatially unstructured component. The proportion of the marginal variance explained by the structured effect is represented by a parameter $\lambda^{t}$ \citep[for more details see][]{Riebler2016}, and $(\sigma_{\xi}^{t})^2$ is the variance parameter of the spatial random effects.

Then, if $\hat{r}_{it}$ represents the posterior median relative risk estimate for the $i$th county with missing value in year $t$, the missing count $Y_{it}$ is substituted with $\hat{Y}_{it}= \text{round}(e_{it}\hat{r}_{it})$ (values rounded to the nearest integer). Subsequently, $\hat{Y}_{it}$ values exceeding 9 are truncated to 9. This adjustment is more meaningful as it accommodates missing data that are always less than or equal 9. Additionally, we also consider the imputation process without truncation. Though final trends do not change too much, relative risk estimates are higher without truncation. A different approach to deal with missing data is to fit spatio-temporal models to the incomplete data and predict the missing counts. As suggested by the reviewers, we also follow this approach, but it produces counts over 9 in more than 50\% of the areas with missing data, leading to a considerable overestimation.

\subsection{Descriptive summary}\label{sec_descrip}

As a preliminary analysis, we computed global and county-level Standardized Mortality Ratios (SMRs) of IHD deaths after the imputation of missing data.}
The map in Figure \ref{fig:SMR_space_time} (top) describes the spatially structured distribution of IHD deaths, representing the global SMRs across space over the entire period computed as $SMR_{i}=\dfrac{\sum_{t=1}^{T} Y_{it}}{\sum_{t=1}^{T} e_{it}}$.  The global SMRs for the counties with counts suppressed in the whole period are not displayed for confidentiality issues.

\begin{figure}[htbp]
	\centering
	\includegraphics[width=0.75\textwidth]{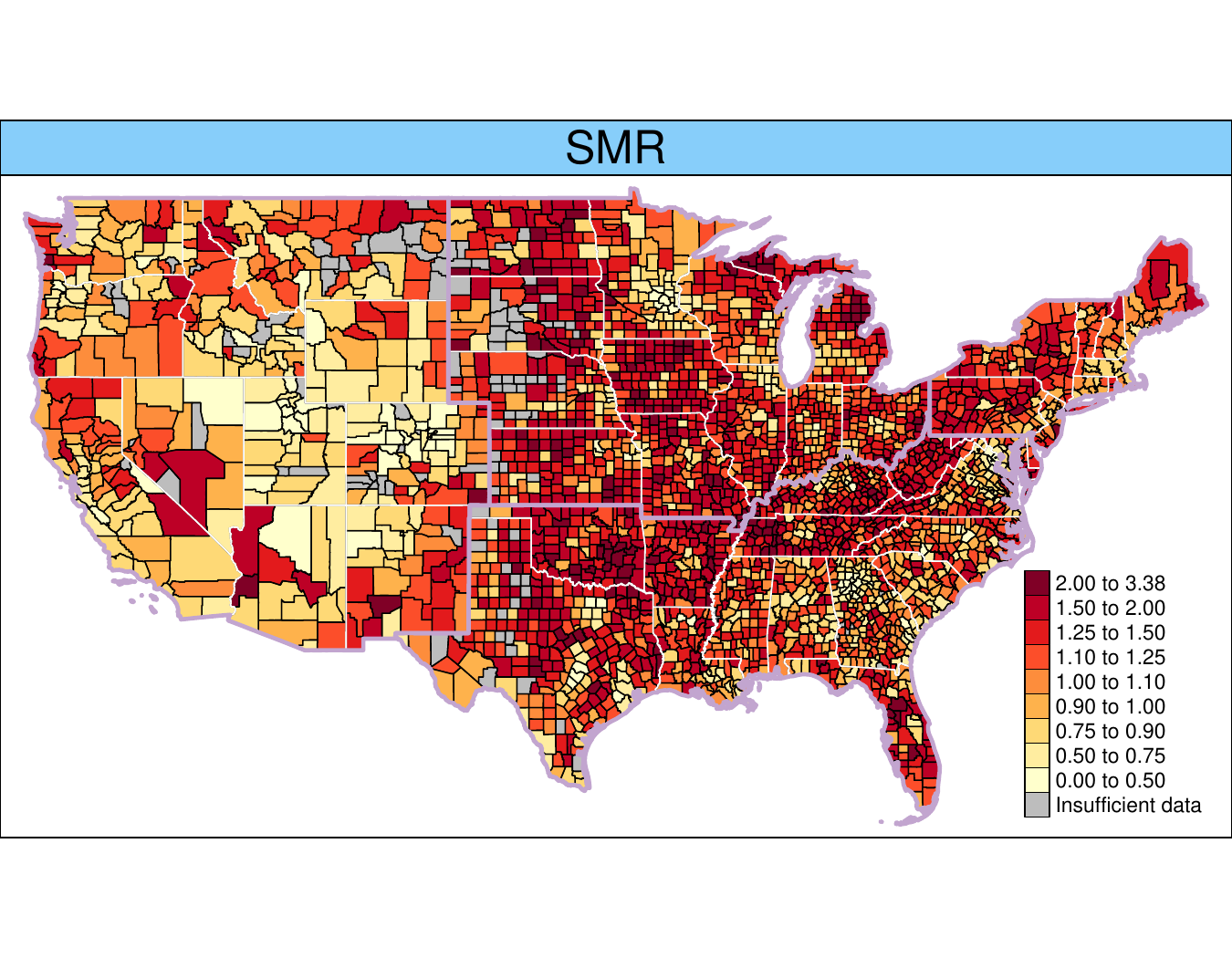}
	\includegraphics[width=0.76\textwidth]{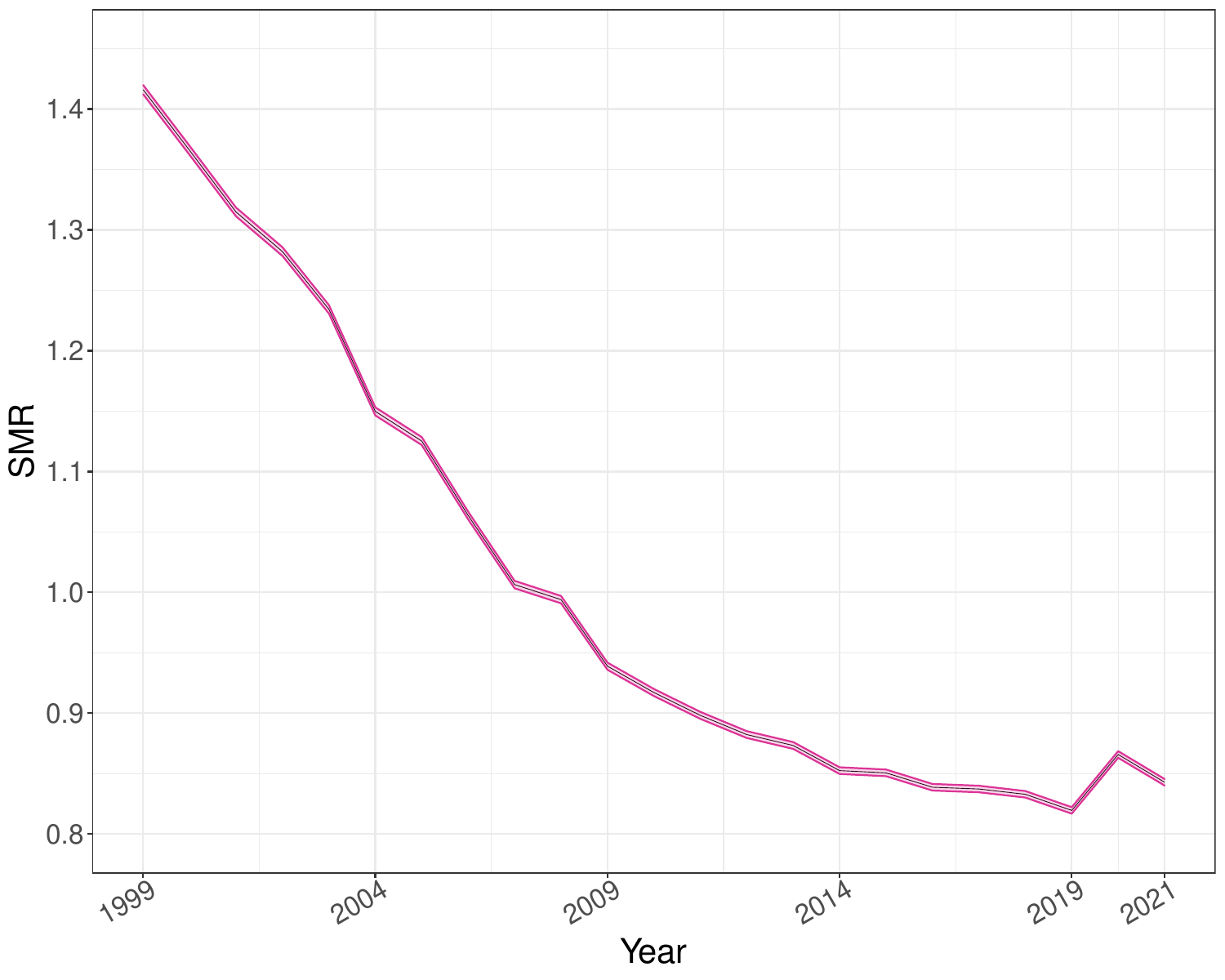}
	\caption{SMRs in space ($SMR_{i}$) over the period 1999-2021 (top). The thickest violet lines are used to limit the four main geographic regions of US, West, Midwest, South and Northeast. The temporal trend of the SMRs ($SMR_{t}$) is shown at the bottom.}
\label{fig:SMR_space_time}
\end{figure}

Generally, counties in the West region exhibit low SMRs. %compared to the Midwest, South and Northeast regions.
Figure \ref{fig:SMR_space_time} (bottom) shows the temporal trend of IHD mortality, which is obtained representing the global SMRs across all US counties at time $t$, calculated as $SMR_{t}=\dfrac{\sum_{i=1}^{S} Y_{it}}{\sum_{i=1}^{S} e_{it}}$, with the corresponding $95\%$ confidence intervals. The temporal trend of IHD shows a consistent decline from 1999 to 2019, although there is some deceleration in the decrease after 2014. Additionally, there is some evidence indicating an increase in IHD deaths after 2019. The number of expected cases have been computed as $e_{it}=N_{it} m$ where $m=\dfrac{\sum_{t=1}^{T}\sum_{i=1}^{S} Y_{it}}{\sum_{t=1}^{T}\sum_{i=1}^{S} N_{it}}$ is the overall rate in the US in the whole period of study. Note that the overall rate $m$ has been computed ignoring the missing data. Nevertheless replacing the missing data with the imputed counts has a negligible effect on the overall rate. Finally it is worth highlighting that the standardized mortality ratios and crude rates are proportional as the expected counts have been obtained using the global rate for the entire region and study period. Hence, in subsequent sections, the estimated relative risks for the different counties can be directly compared.

Though the global spatial and temporal SMRs are rather stable (they are computed over the whole period and over all the counties respectively), a close inspection to the data in each county and time period reveals high variability. Table \ref{tab:descrip} displays the minimum and maximum number of observed counts ($Y_{it}$) in 1999, 2010 and 2021, together with the corresponding number of expected cases ($e_{it}$), the standardized mortality ratio $SMR_{it}$, the variance of the $SMR_{it}$ (var$(SMR_{it})$) and the coefficient of variation of the $SMR_{it}$ (cv$(SMR_{it})$). Clearly, the SMR for areas with few cases is highly variable with unacceptably large coefficients of variation.

\begin{table}[ht]
    \centering
    \caption{Descriptive statistics. Minimum and maximum number of observed counts ($Y_{it}$) in 1999, 2010 and 2021, the corresponding number of expected cases ($e_{it}$), the standardized mortality ratio $SMR_{it}$, the variance of the $SMR_{it}$ (var$(SMR_{it})$) and the coefficient of variation of the $SMR_{it}$ (cv$(SMR_{it})$).}
    \begin{tabular}{ccccccc}
        \hline
    Year & Min/Max & $Y_{it}$ & $e_{it}$ & $SMR_{it}$ & var$(SMR_{it})$ & cv$(SMR_{it})$ \\
   \hline
    \multirow{2}[2]{*}{1999} & Min & 1     & 1.3561     & 0.7374 & 0.5438 & 1.0000 \\
                             & Max & 16943 & 12695.9827 & 1.3345 & 0.0001 & 0.0077 \\
    \hline
    \multirow{2}[2]{*}{2010} & Min & 1 & 1.5027 & 0.6655 & 0.4428 & 1.0000 \\
    & Max & 11860 & 13208.9656 & 0.8979 & 0.0001 & 0.0092 \\
    \hline
    \multirow{2}[2]{*}{2021} & Min & 1 & 1.5323 & 0.6526 & 0.4259 & 1.0000 \\
    & Max & 11335 & 13223.6818 & 0.8572 & 0.0001 & 0.0094 \\
    \hline   \end{tabular}
   \label{tab:descrip}
\end{table}

\begin{figure}[htbp]
	\centering
	\includegraphics[width=1\textwidth]{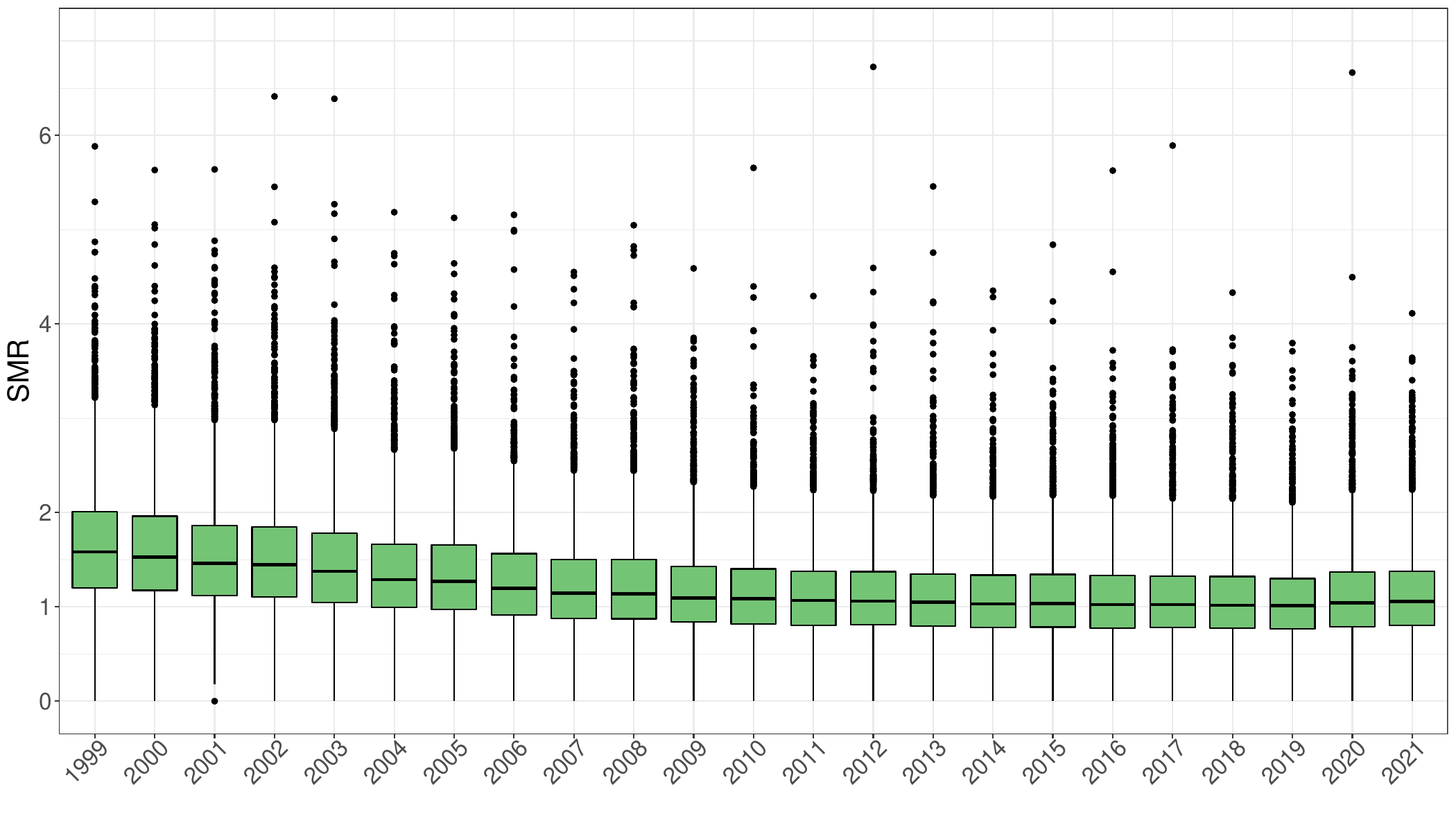}
	\caption{Boxplots of county spatio-temporal SMRs grouped by year.}
	\label{fig:SMR_boxplot}
\end{figure}

Finally, Figure \ref{fig:SMR_boxplot} displays the boxplots of the spatio-temporal SMRs ($SMR_{it}$) for all counties grouped by years. The decreasing trend in time is observed in the median of the boxplots, but the large variability in the SMRs becomes evident, highlighting the necessity for spatio-temporal models to provide risk estimates at the county level.

\subsection{Spatio-temporal models}

Spatio-temporal disease mapping models are widely used to describe the geographical pattern of a disease and its evolution in time. In this study, we will apply spatio-temporal models to understand the temporal evolution of IHD mortality risks at a county level. Various proposals are available for the spatial, temporal, and spatio-temporal interactions components of the model.  Interactions are an advisable feature to reflect different trajectories of decline and slowed decline between different regions and counties. We also need a flexible model to account for non-linear features in the overall mortality trend.

In disease mapping, conditional autoregressive (CAR) \citep{Besag1974} and random walk (RW) priors are widely adopted for spatial and temporal random effects respectively, and provide that flexibility. However, alternatives based on P-splines can be used instead \citep{Goicoaetal2012, Ugarte2012AN}. To capture localized trend variation,  spatio-temporal interactions can be based on Markov
random fields \citep{Knorr-Held2000} or P-splines \citep{Ugarte2010}.

The spatio-temporal model adopted is the following one
\begin{eqnarray}\label{eq:spatio_temporal _model}
   Y_{it}\arrowvert r_{it} &\sim& Poisson(e_{it}r_{it}) \\\nonumber
	\log r_{it}&=&\alpha_0+\xi_{i}+\gamma_{t} + \delta_{it},
\end{eqnarray}
where $\alpha_{0}$ can be interpreted as an overall risk level and $\boldsymbol{\xi}=(\xi_1, \xi_2, \dots, \xi_{S})^{'}$ is the vector of spatial random effects that follows an ICAR prior, $\boldsymbol{\xi} \sim N(\boldsymbol{0}, \sigma^2_{\xi}\Q_{\xi}^{-})$, or BYM2 prior as in model (\ref{eq:spatial_model}).
The vector of temporal random effects $\boldsymbol{\gamma}=(\gamma_1, \gamma_2, \dots, \gamma_{T})^{'}$ is assumed to follow a first order random walk (RW1). Then, $\boldsymbol{\gamma} \sim N(\mathbf{0}, \sigma_{\gamma}^2\Q_{\gamma}^{-})$, where $\Q_{\gamma}$ is a the precision matrix of a RW1 \citep[for more details see][page 95]{Rue2005}, and $\sigma_{\gamma}^{2}$ is the variance parameter.
%$\boldsymbol{\gamma}=(\gamma_1, \gamma_2, \dots, \gamma_{T})^{'}$ is the vector of temporal random effects for which a first order random walk (RW1) prior is assumed. Then, $\boldsymbol{\gamma} \sim N(\mathbf{0}, \sigma_{\gamma}^2\Q_{\gamma}^{-})$ where $\Q_{\gamma}$ is a ``neighbourhood'' matrix in time where each time point (except the first and the last one) has two neighbours, the preceding and the subsequent one \citep[for more details see][page 95]{Rue2005}, and $\sigma_{\gamma}^{2}$ is the variance parameter.
Here the symbol $-$ indicates the Moore-Penrose generalized inverse. The vector of spatio-temporal random effects $\boldsymbol{\delta}=(\delta_{11}, \dots, \delta_{S1}, \dots, \delta_{1T}, \dots, \delta_{ST})^{'}$ captures the specificities of the counties in each time point. The interaction terms are assumed to be normally distributed as $\boldsymbol{\delta}\sim N(\mathbf{0}, \sigma_{\delta}^2\OOmega_{\delta})$ where $\sigma_{\delta}^{2}$ is the variance parameter. Here, we will define the covariance matrix $\OOmega_{\delta}$ according to the four types of space-time interactions introduced by \cite{Knorr-Held2000}: in Type I interaction all the elements of $\boldsymbol{\delta}$ are independent; in Type II the elements are structured in time but not in space; in Type III interaction the terms are structured in space but not in time; finally in Type IV interaction the terms are structured in space and time.  %{\color{red} The latter may produce better fit but is highly structured, and may be judged not needed by fit measures which penalize complexity.}

To address the computational challenges  that involve fitting the classical global spatio-temporal models incorporating Type II, Type III, and Type IV interactions due to the large number of counties, we opt for a partitioned modeling approach implemented in the \verb"bigDM"  package \citep{OrozcoAcosta2023}. The \texttt{R} package \verb"bigDM" is tailored for fitting Bayesian disease mapping models with high-dimensional data using a divide-and-conquer strategy. This method involves applying spatio-temporal models across multiple partitions within the region of interest and accelerates computations. Moreover, these localized models exhibit remarkable adaptability in capturing spatial and temporal heterogeneity, employing distinct precision parameters in different subdivisions \citep{OROZCOACOSTA2021}. In contrast, the global model, which employs the same smoothing parameters across all regions, may be questionable when dealing with a large number of small areas. We have also scaled the precision matrix of the spatial priors as discussed in \cite{Riebler2016}. While this does not change the relative risk estimates, it allows for a direct comparison of the spatial precision/variance parameters across the different partitions, indicating varying extents of spatial smoothing.

\subsection{Spatio-temporal model fitting}

 The \lq\lq divide and conquer'' approach requires a partition of the study region into smaller subdomains \citep[Section 3.1]{OROZCOACOSTA2021}, and
here we take the US states as the natural subdivision. However, three states have an insufficient number of counties to allow for effective partitioning. Namely, Delaware consists of three counties, the District of Columbia has a single county, and Rhode Island encompasses five counties. Hence, the District of Columbia, Maryland, and Delaware are treated as a single state, and Rhode Island is combined with Connecticut. After this adjustment, Connecticut consists of a total of 13 counties and the combined state of District of Columbia, Maryland, and Delaware has 28 counties. The \lq\lq divide and conquer'' approach used here is robust to the partition chosen \citep[see][]{OROZCOACOSTA2021}, but in general it is advisable to reach a trade-off between the number of subdivisions and the number of counties within each subdivision. Furthermore, it is interesting to use large administrative divisions (as states) because the results for each partition could be more useful for health authorities.
To mitigate border effects in the risks estimates, the modeling approach adds $k$-order neighbours to the counties located in the border of each subregion. In this case we have chosen first-order neighbours ($k=1$).

A total of 8 sets of partitioned models have been fitted to the IHD mortality data with imputed counts combining the two spatial priors (ICAR and BYM2) and the four types of interactions. To select the most appropriate model, Table \ref{table:goodnessfit} provides the mean deviance ($\bar{D}$), the effective number of parameters ($p_{D}$), the Deviance Information Criterion (DIC) and the Watanabe-Akaike Information Criterion (WAIC). We note that rather than presenting raw values for the DIC and WAIC, the displayed tables represent the difference between the values of the DIC and WAIC with respect to the minimum value. Hence, the value 0 indicates the best model.
 The spatio-temporal model with the ICAR spatial prior and Type II interaction exhibits the lowest values of the DIC and WAIC criteria.

\begin{table}[ht]
	\centering
	\caption{The mean deviance ($\bar{D}$), effective number of parameters ($p_{D}$), DIC and WAIC (difference with respect to the minimum value) for the partitioned spatio-temporal models fitted to the IHD mortality data with imputed values.}
	\begin{tabular}{clcccc}
		\hline
		& & $\bar{D}$ & $p_{D}$ & DIC & WAIC \\
		\hline
		\multirow{4}[2]{*}{ICAR} & TypeI & 470549.7564 & 31018.9704 & 12227.4389 & 9997.6327 \\
		& TypeII & 470630.9527 & 18710.3352 & 0.0000 & 0.0000 \\
		& TypeIII & 476188.1729 & 26080.9460 & 12927.8310 & 14382.4661 \\
		& TypeIV & 473448.4520 & 16759.8658 & 867.0299 & 1865.6019 \\
		\hline
		\multirow{4}[2]{*}{BYM2} & TypeI & 470600.7076 & 31021.2286 & 12280.6483 & 10101.1408 \\
		& TypeII & 470645.9054 & 18726.7258 & 31.3433 & 48.9572 \\
		& TypeIII & 476225.8719 & 26102.6948 & 12987.2788 & 14483.4961 \\
		& TypeIV & 473457.6642 & 16770.4608 & 886.8371 & 1884.9096 \\
		\hline
	\end{tabular}
	\label{table:goodnessfit}
\end{table}

Additionally, to compare the partitioned and classical global spatio-temporal models, Table C.1 in Supplementary Material C presents the model selection criteria for both the partitioned and global models with the ICAR prior. Global models with the BYM2 prior and the four interactions are not fitted due to computational difficulties. This means that these models cannot be fitted due to large number of areas and time points, even in a computer cluster with 80 CPUs Intel(R) Xeon(R) Silver 4316@ 2.30GHz and 250 GB RAM. According to Table C.1, the partitioned models outperformed the global models. Therefore, all the subsequent results are associated with the partitioned spatio-temporal model with the ICAR spatial prior, a RW1 temporal prior and a Type II spatio-temporal interaction. This model took about 13.8 minutes on a desktop computer with 3.00 GHz Intel(R) Core(TM) i5-9500 CPU processor and 20GB of RAM. % El paper de Quick Estimating County-Level Mortality Rates Using Highly Censored Data From CDC WONDER analiza 6 grupos de edad en 3109 counties en 1980. Esto dice que en R cuesta 4 horas y en winBugs 20 minutos en su portátil.

All models in the paper were fitted using \texttt{R} version 4.2.1 and the \texttt{R-INLA} package \citep{LindRue2015} version 22.12.16 (dated 2022-12-23) which considers Laplace approximation method with a low-rank Variational Bayes correction to the posterior mean \citep{VanNiekerk2023a, VanNiekerk2023b}. For partitioned modelling, we utilized version 0.5.3 of the \verb"bigDM" package. The full code and data to reproduce the results in this paper are available at {\color{blue}\url{https://github.com/spatialstatisticsupna/IHD_ST_patterns}}.

%%%%%%%%%%%
% RESULTS %
%%%%%%%%%%%

\section{Results}

\begin{figure}[h!]
	\centering
\includegraphics[width=1\textwidth]{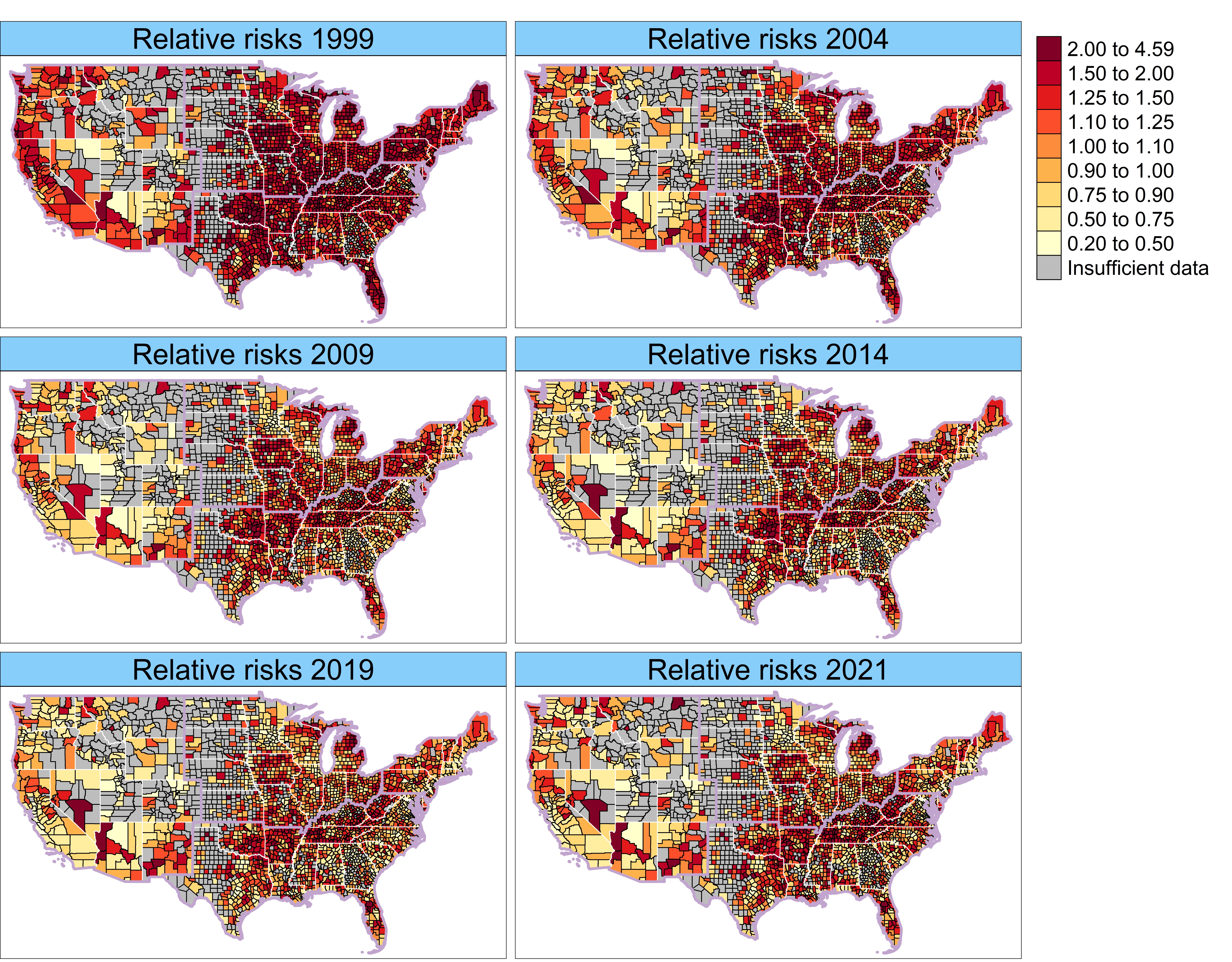}
	\caption{Posterior median estimates of the relative risks, $\hat{r}_{it}=exp(\alpha_0+\xi_{i}+\gamma_{t}+\delta_{it})$, for the years 1999, 2004, 2009, 2014, 2019 and 2021 derived from the partitioned spatio-temporal model with ICAR prior and Type II interaction.  The relative risks of counties with missing counts are not displayed.}
	\label{fig:RR_medians}
\end{figure}

Figure \ref{fig:RR_medians} displays the posterior median estimates of relative risks for the selected years 1999, 2004, 2009, 2014, 2019, and 2021.
A noticeable decreasing trend in relative risks is evident from 1991 to 2021, as illustrated by the maps gradually lightening over time.
%A noticeable decreasing trend of the relative risks is evident from 1991 to 2021, depicted by the maps gradually lightening over time.
In more detail, Table D.1 in the Supplementary Material D provides the posterior medians and $95\%$ credible intervals for the relative risks of the counties containing the capitals of 48 US states (excluding Hawaii and Alaska). For each county, the highest relative risk value across all the analyzed years is highlighted in red.
This table supports the pattern observed in Figure 4, where the highest risk estimates are predominantly seen in 1999 and then decrease throughout the study period.

%This table supports the observed pattern in Figure \ref{fig:RR_medians}, wherein the highest risk estimates are predominantly observed in 1999 and then exhibit a decrease %throughout the study period.

\begin{figure}[h!]
	\centering
	\includegraphics[width=0.95\textwidth]{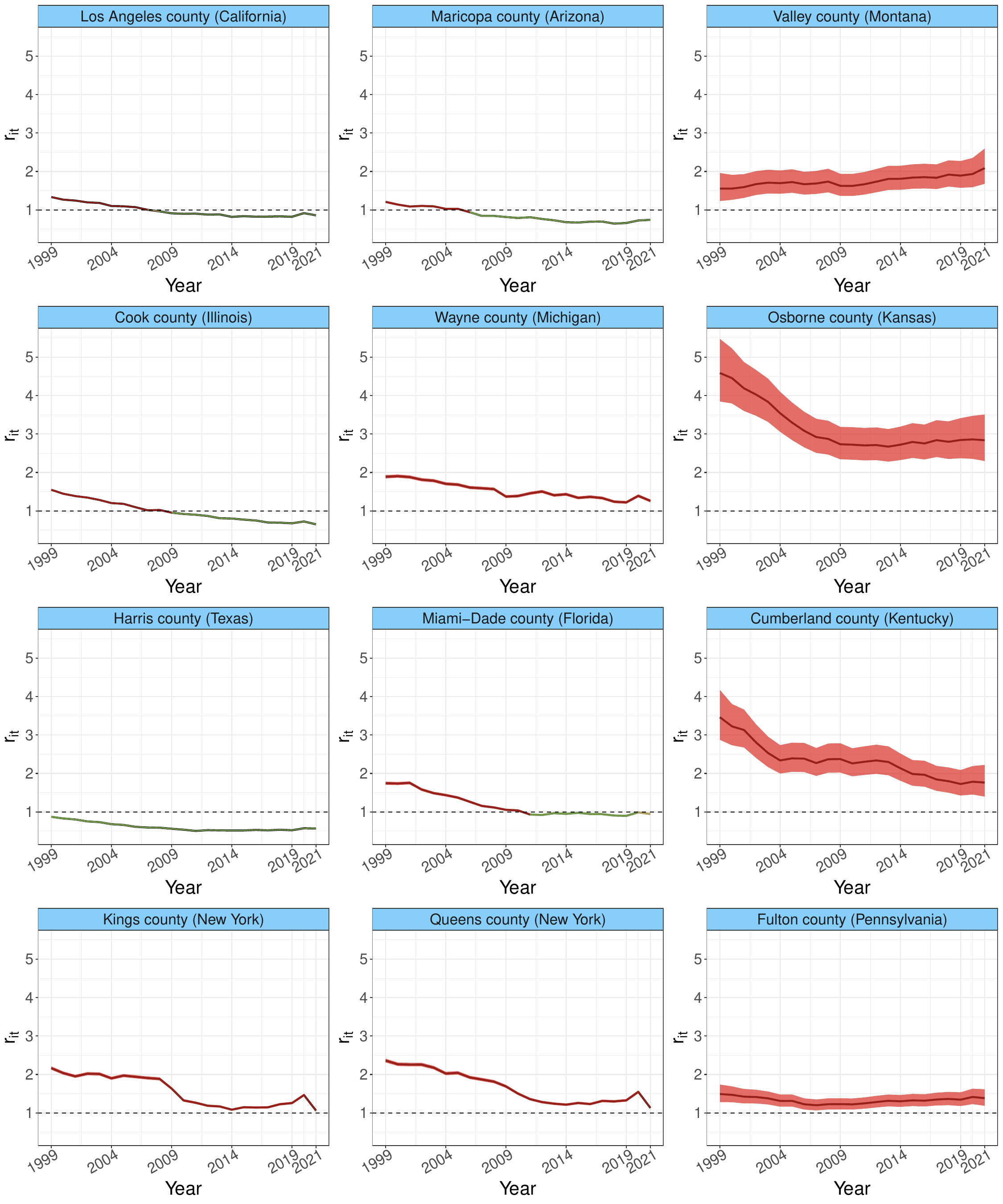}
	\caption{Temporal evolution of IHD mortality risks ($\hat{r}_{it}=\text{exp}(\alpha_0+\xi_{i}+\gamma_{t}+\delta_{it})$) and their $95\%$ credible intervals, for most populated counties (first and second columns) as well as the less populated county  without suppressed counts (third column) within the West (top row), Midwest (second row), South (third row) and Northeast (bottom row) regions. The red color indicates a high posterior probability of relative risks being greater than one.}
	\label{fig:RR_temporal_trend}
\end{figure}

Figure~\ref{fig:RR_temporal_trend} illustrates the temporal evolution of estimated IHD mortality risks, including their $95\%$ credible intervals, for some counties. The first and second columns correspond to the most populated counties in the West (top row), Midwest (second row), South (third row) and Northeast (bottom row) of US whereas the last column refers to the less populated counties  with more than 9 cases in each year of the period. The red color in the credible intervals indicates a high posterior probability that the  relative risks are greater than one. The objective is to assess whether disparities exist between these counties and the overall temporal trend of IHD mortality in the US, as well as to identify potential change points in the trends.

%The objective is to assess if there exist disparities between these counties and the overall temporal trend of IHD mortality in the US and to identify potential change points in %the trends.

Overall, in a majority of counties in Figure~\ref{fig:RR_temporal_trend}, there is a deceleration in risk reduction after 2009-2014 and an increase in risk levels after 2019. This is consistent with previous research in \cite{Mehta2020}, which suggest a slowdown in the decline of mortality risks after 2010. Moreover, in line with \cite{Kulshreshtha2014}, there are disparities between the temporal trends across counties in different geographic regions of US. This contrast is particularly pronounced in the counties of New York (Kings and Queens counties in Northeast) where the trends markedly differ from those counties in the West, Midwest, and South as well as to the overall US trend.
Furthermore, mortality risks in New York counties consistently exceed one throughout the entire study period. Disparities in trends are also evident in other counties, such as Wayne County (Michigan) in the Midwest and Harris County (Texas) in the South. Additional disparities are observed in sparsely populated counties. Notably, Valley County (Montana) in the West exhibits a rising trend, while Fulton County (Pennsylvania) in the Northeast maintains a consistent pattern.
Overall we notice an excess of risk in low populated counties (right column). Despite the wider 95\% credible intervals, they are considered reliable as their width is smaller than that of the risk point estimates (posterior medians) \citep[see for example][]{Vaughan2022a,Vaughan2022b}. To inspect the IHD mortality risk trends in each county, an interactive map is available at  {\color{blue}\url{https://emi-sstcdapp.unavarra.es/IHD_leaflet_map.html}}.
%{\color{blue}\href{https://emi-sstcdapp.unavarra.es/IHD_leaflet_map.html}{here}}.
Recall that we can directly compare the relative risks of the different counties because we have used a global rate to compute the expected cases and hence, rates and relative risks are proportional.

%\clearpage

To inspect the results in more detail, Table D.2 in the Supplementary Material D displays the relative risk estimates for counties whose lower boundary of the $95\%$ credible interval in 2021 exceeds 1 (high risk areas). Similarly, Table D.3 presents the relative risk estimates for counties where the upper boundary of the $95\%$ credible interval in 2021 is below 1 (low risk areas). Among all the identified high- and low-risk areas, Tables D.2 and D.3 provide the relative risk values for each year pertaining to the 10 counties with the highest and lowest populations in 2021, excluding those with missing counts. The relative risks generally exhibit a declining trend until 2014, followed by a flattening or increase observed thereafter.
% Among all the identified high and low risk areas, Tables D.2 and D.3 provide the relative risk values for each year pertaining to the 10 counties with the highest and lowest %populations in the year 2021 without missing counts in the years represented. The relative risks generally exhibit a declining trend until 2014, with a subsequent flattening or %increase observed afterward.

\begin{figure}[h]
	\centering
	\includegraphics[width=0.65\textwidth]{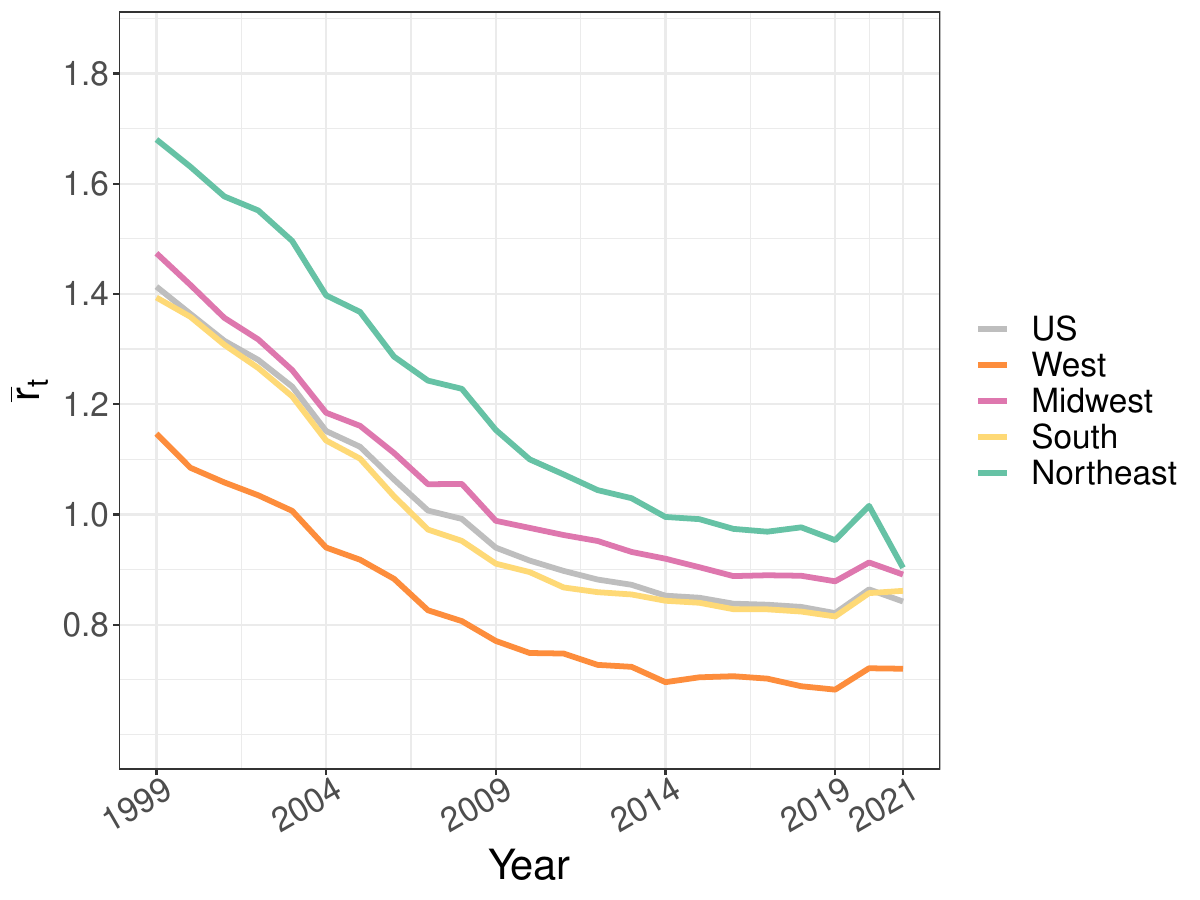}
	\caption{The mean of IHD mortality risks of the counties in West (orange), Midwest (pink), South (yellow) and Northeast (green) regions during the period 1999-2021. Temporal evolution of the mean IHD mortality risks of all the counties in the US is represented in grey color.}
	\label{fig:RR_temporal_trend_4regions}
\end{figure}

To visualize disparities in the temporal evolution of IHD deaths among the four big geographic regions, West, Midwest, South, and Northeast, we represent the mean temporal risk trends (Figure \ref{fig:RR_temporal_trend_4regions}) of the counties within each region, weighted by the county population. The weighted average temporal trend of the Midwest and, more specifically of the South, closely mirror the overall US trend. The weighted average trend for the West is clearly below the rest of regions, and the mean trend of the Northeast region is above all of them. Note that, unlike the other three regions, in the Northeast there are more urban than rural counties. Across all geographic regions, a slowdown in the decrease is observed after 2009 and a flattening in the trend is observed after 2014. A peak is observed in 2020, but caution is recommended because it corresponds to the first year of the COVID-19 pandemic. As an example, \cite{ROTH2022} indicates that increases in cardiovascular deaths in 2020 might be due to misclassification of COVID-19 deaths, though more research is needed to ascertain to what extent the pandemic may explain this increment.

\begin{figure}[h]
	\centering
	\includegraphics[width=1\textwidth]{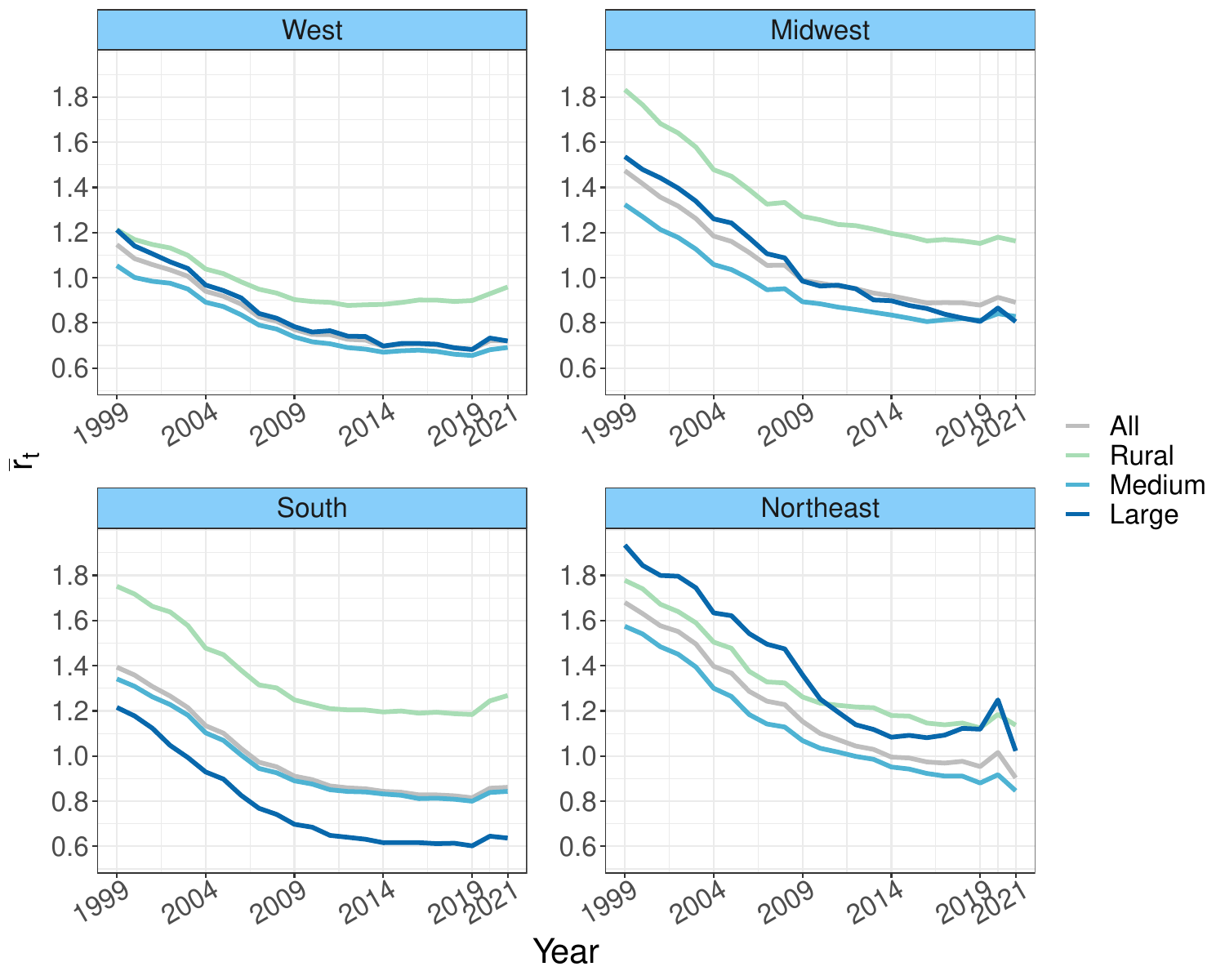}
	\caption{Mean of IHD mortality risks in Rural (green), Medium (light blue) and Large (dark blue) metros classifying by the geographic region (West, Midwest, South and Northeast) in the period 1999-2021. In grey color the temporal trend of IHD risks in each geographic region without classifying by demography is represented.}
	\label{fig:RR_demographic_groups_4regions}
\end{figure}

Additionally, the temporal trends of IHD mortality risks are represented categorized by demographic groups. Figure \ref{fig:RR_demographic_groups_4regions} represents the weighted average trends for urban and rural areas classified by geographic region. In the West, Midwest, and South regions, the weighted average IHD mortality risk trends are greater in rural than in urban areas. Comparing urban areas, large metros have higher average trends than median metros in the West and Midwest, whereas in the South is just the opposite. The pattern differs in the Northeast region. In this area, large metros have higher average risks than rural areas until 2009. From this year onwards, the average risks in rural areas is the highest with the exception of 2020 (the starting of the COVID-19 pandemic), where a peak is observed in the trend of large metros. Regarding median metros, they present the lowest average risk trend during the study period. The higher risks in rural areas across the West, Midwest, and South regions could be influenced by the number of imputed values in counties with very low population, potentially leading to some overestimation in these counties. To investigate this, we have compared the weighted average temporal trends for rural areas within each of these large regions, as estimated by our model, with the temporal Standardized Mortality Ratios (SMRs) for rural areas across counties that have observed counts (excluding imputed areas). At this level of aggregation, the SMR is a reliable measure. We do not observe an overestimation of the temporal trends obtained with our model compared to the SMRs. Therefore, it appears that the imputation procedure does not impact the results, at least at some aggregated level. We think that the potential overestimation might be due to the smoothing process itself. Spatial and spatio-temporal smoothing borrow strength from neighbouring areas with observed counts that are greater than 10, which may push up the risk of the areas with missing counts.
We have also fitted a spatio-temporal model to the original data with missing information and we have predicted the censored counts. The results  (not shown here to save space) show that this latter procedure overestimates the missing counts more than our method. More precisely, the spatio-temporal model fitted to the data with missing information predicts counts over 9 in more than 50\% of the areas with missing data, whereas our approach predicts counts over 9 in 25\% of the areas with censored counts.

%{\color{red}The elevated risks in rural areas of the West, Midwest, and South could be influenced by the high number of imputed values in counties with very low populations, potentially leading to overestimations in these counties. To assess this, we have compared the temporal trends for rural areas within each large region estimated with our model with the temporal SMR for rural areas across counties with observed counts (imputed areas are excluded). At this level of aggregation, the SMR is a reliable measure. We do not observed an overestimation of the temporal trends of our model compared to the SMRs. Therefore, it appears that the imputation procedure does not impact the results, at least at the four regions level, and the temporal trends for rural areas shown in Figure \ref{fig:RR_demographic_groups_4regions} are reliable.}

In summary, the average risk trend in the US decreases until 2014 and then remains rather flat. However, an increase is observed in 2020, the starting of the COVID-19 pandemic. This general behaviour is observed in the four big geographic regions. However, when computing average risk trends by population subgroups and big regions, differences emerge. In particular, the increase in trend from 2019 onwards is not so evident in the Midwest and Northeast. Finally, when examining the risk trends by county
(see {\color{blue}\url{https://emi-sstcdapp.unavarra.es/IHD_leaflet_map.html}}), disparities become evident. Some counties display decreasing trends, while others show increasing trends, such as  Valley county in Montana or Menominee county in Michigan.

Finally, we would like to emphasize that in addition to computational advantages, the \lq\lq divide and conquer'' approach offers a more adaptive smoothing as we estimate different precision/variance parameters for each partition. In our case study, the posterior estimates of the spatial standard deviations in each subregion range from the minimum value 0.212 to the maximum value 0.519. As we have scaled the spatial precision matrix, these values are comparable and reflect varying degrees of spatial smoothing. This variation suggests that using a single spatial variance parameter for the entire region (global model) may not be appropriate.

%%%%%%%%%%%%%%%%
%% DISCUSSION %%
%%%%%%%%%%%%%%%%
\clearpage
\section{Discussion}

In this paper, we use an efficient procedure to screen spatio-temporal patterns of IHD mortality in US counties. Additionally, we aim to identify counties that demonstrate disparities in their temporal trends compared to the overall trends in the US.% and to visually detect change points in the time trends}.
In our study, we utilize publicly available IHD censored data from CDC WONDER, though access to the complete NCHS registries may be possible subject to specific terms of use. The IHD data from CDC WONDER contain suppressed information to maintain confidentiality when the count in a given county and year is less than 10. Consequently, we face two primary challenges: handling the missing information and fitting Bayesian spatio-temporal models much faster than previous proposals in the literature. %to a large dataset.
To address the censored data, our approach involves fitting spatial models for each year of the study period and replacing the missing information with predicted counts derived from these models. Additionally, when the predicted counts for the missing values exceed 9, we truncate them, as suppressed data refers to counts less than or equal to 9. Following the reviewers' recommendations, we also fitted a spatio-temporal model with the missing values and used it to predict counts in areas with censored counts. However, this approach overestimated the missing counts more than our procedure. More precisely, it produces counts over 9 in more than 50\% of the areas with missing data. Regarding problems with Bayesian inference on large datasets, we adopt a scalable Bayesian modeling approach employing a \lq\lq divide and conquer'' strategy. This strategy entails partitioning the study domain into smaller subregions and fitting spatio-temporal models within those subdomains. Besides providing computational advantages, this scalable approach demonstrates flexibility by allowing the adjustment of smoothing parameters tailored to individual partitions, thereby outperforming global models that rely on a single smoothing parameter for the entire region. In extensive areas divided into a large number of regions (as it is the case of US), a global model somehow assumes stationary in the data that may not be reasonable. The scalable approach produces a kind of spatial adaptive smoothing that may better capture local heterogeneities, as the different estimates of the spatial standard deviations in the partitions show. Furthermore, the method effectively mitigates border effects by incorporating k-order neighbours within each subdomain.

Analysis of the IHD data yields some interesting findings, some of which are consistent with previous research \citep{Cooper2000, Kulshreshtha2014, Mehta2020, Vaughan2020}. The overall temporal trend of mortality risk in the US displays a noticeable decline until 2019,  but a deceleration in risk reduction is observed after 2009. An increase in risk becomes apparent after 2019, though further analyses using subsequent years data are necessary to confirm  if this increase is persistent or it may be related to the effects of COVID-19 pandemic on mortality. Consequently, our analysis suggests 2009 and 2019 as potential change points in the overall temporal trend of IHD mortality risk.
As noted in \cite{Vaughan2017}, while the overarching trend depicted a decrease, a majority of counties and age groups experienced increases in heart diseases from 2010 to 2015, unveiling inequalities in IHD mortality. Other disparities among counties, such as the timing of recent increases in heart diseases \citep{Vaughan2022a}, or increasing of coronary heart diseases (CHD) in non-white adult population  in some counties in Mississippi, Oklahoma, Texas and New Mexico \citep{Vaughan2020} have been documented. In our study, some counties like Los Angeles, Maricopa, or Wayne, mirror the global temporal pattern, whereas others such as Kings or Queens exhibit distinct trends with varying periods of decline and rise.  Moreover, some counties like Menominee in Michigan or Valley in Montana show increasing trends. Hence, the decline in the national trend does not necessarily manifest in local trends. Further research is needed to identify potential risk factors related to these discrepancies between the trends in some counties and the overall regional trend. Some research \citep[see, for example][]{Wilmot2015, Vaughan2017, Vaughan2018} points towards the increase of the prevalence of obesity and diabetes.

%{\color{red}Further research is needed to identify potential risk factors related to these discrepancies observed in some counties compared to the overall regional trend. }

%The analysis of IHD data provides some interesting findings and corroborates those in previous research \textcolor{blue}{\citep{Cooper2000, Kulshreshtha2014, Mehta2020, Vaughan2020}}. The overall temporal trend of mortality risk in the US displays a noticeable decline until 2019, followed by a deceleration in risk reduction after 2009. An increase in risk becomes apparent after 2019, though further analyses using subsequent years data are necessary to confirm  if this increase is persistent or it is just a random fluctuation. Consequently, our analysis suggests 2009 and 2019 as potential change points in the overall temporal trend of IHD mortality risk.
%\textcolor{blue}{As noted in \cite{Vaughan2017},} while the overarching trend depicts a decrease, county-specific temporal trends highlight disparities, unveiling inequalities in IHD mortality. Some counties, like Los Angeles, Maricopa, or Wayne, mirror the global temporal pattern. However, others such as Kings or Queens exhibit distinct trends with varying periods of decline and rise.  Moreover, some counties like Menominee in Michigan or \textcolor{blue}{Valley} in Montana show increasing trends. Hence, the shift in trends in the global pattern does not necessarily manifest in local trends.

Examining the temporal trends of IHD mortality risks specific to each geographic region is helpful in gaining a comprehensive overview of the trends. The mortality risks due to IHD are relatively low in the West region compared to the overall temporal trend, but they remain concerning in the other regions, particularly in the Northeast. Moreover, after classifying the counties as urban and rural areas, relevant discrepancies are observed among these demographic groups in the West, Midwest, and South regions, with higher IHD  weighted average mortality risks in rural areas.  On the contrary, in the Northeast the large metro group has the highest weighted average risks until 2009, and from then on, the average trend for rural areas is above the rest. These findings support the conclusion of  \cite{Kulshreshtha2014} which emphasizes the existence of disparities in certain geographic regions and demographic groups.

Our study may have some limitations. Firstly, counts less than 10  are suppressed due to confidentiality reasons, and thus, they have been imputed. Approximately 10\% of the counties (310) exhibit missing values across nearly all the years. While we truncated predicted counts above 9 during the imputation process, our results may be sensitive to this method. Since missing values consistently fall below 10 and observed counts are at least 10, there is potential for some inflation in the imputed values due to spatial borrowing of strength. However, this potential inflation might be due to the smoothing process and the spatial priors themselves rather than to the imputation approach. Difficulties in estimating rates for rural areas and minority populations with CDC WONDER data has already been noticed \citep{Quick2019}.
Secondly, confirming the rise in mortality risk after 2019 requires further analysis. With available data only until 2021, a two-year span might not suffice to distinguish between a persistent increase or a transient effect of the COVID19 pandemic. Finally, our data set lacks the granularity to study temporal IHD mortality trends among other demographic subgroups, such as black or white populations as in the work by \cite{Vaughan2019, Vaughan2020, Vaughan2021}. Additionally, examining IHD mortality risk at the county level across different population subgroups poses a challenge as the suppressed data issue would be enhanced if censored CDC WONDER data are used. For those studies, uncensored data should be advisable.

 We would also like to emphasize the strengths of our method. Specifically, its computational advantages, straightforward and intuitive interpretation of spatio-temporal trends, and its ability to capture local heterogeneities via distinct smoothing parameters for the partitions. We are currently working on a spatio-temporal extension of the spatial multivariate \lq\lq divide and conquer'' approach \citep{vicentehigh2023} to explore further analyses. Finally, there is still room for further research. Rather than imputing censored counts, an interesting alternative would be to extend the "divide and conquer" approach to spatio-temporal models that directly incorporate censored data, such as the model proposed by \cite{Valeriano2021}.

\section*{Acknowledgements}
This work has been supported
by Project PID2020-113125RB-I00/MCIN/AEI/10.13039/501100011033 and by Project UNEDPAM/PI/PR24/05A.

%%---------------------------------------------------------------------------------------
%% REFERENCES
%%---------------------------------------------------------------------------------------
%\clearpage
%\bibliographystyle{apalike}
%\bibliographystyle{munich}

%\bibliographystyle{apalike-ejor}
%\bibliography{references}
%\nocite{}

\end{document}